\documentclass[twocolumn,pra,letterpaper,superscriptaddress]{revtex4-1}
\usepackage{latexsym,amsmath,amssymb,amsfonts,graphicx,color,amsthm,qcircuit}
\usepackage[dvipsnames]{xcolor}
\usepackage{mathrsfs}
\usepackage{multirow}
\usepackage{hyperref}
\usepackage{bm}

\newcommand{\id}{\bm{1}}
\newcommand{\tr}{\mathrm{Tr}}
\newcommand{\upi}{\mathrm{i}}
\newcommand{\upe}{\mathrm{e}}

\newcommand{\cB}{\mathcal{B}}
\newcommand{\cC}{\mathcal{C}}

\newcommand{\cE}{\mathcal{E}}

\newcommand{\cH}{\mathcal{H}}

\newcommand{\cM}{\mathcal{M}}

\newcommand{\cP}{\mathcal{P}}

\newcommand{\cR}{\mathcal{R}}

\newcommand{\cW}{\mathcal{W}}

\usepackage[normalem]{ulem}

\begin{document}

\title{Finding good quantum codes using the Cartan form}
\author{Akshaya Jayashankar}
\thanks{Both authors contributed equally.}
\affiliation{Department of Physics, Indian Institute of Technology Madras, Chennai, India}
\author{Anjala M Babu}
\thanks{Both authors contributed equally.}
\affiliation{Department of Physics, Southern Illinois Univerity, Carbondale, Illinois}
\author{Hui Khoon Ng}
\affiliation{Yale-NUS College, Singapore}
\affiliation{Centre for Quantum Technologies, National University of Singapore{, Singapore}}
\affiliation{MajuLab, International Joint Research Unit UMI 3654, CNRS, Universit{\'e} C{\^o}te d'Azur,
Sorbonne Universit{\'e}, National University of Singapore, Nanyang Technological University, 
Singapore}
\author{Prabha Mandayam}
\affiliation{Department of Physics, Indian Institute of Technology Madras, Chennai, India}

\date{\today}

\begin{abstract}
We present a simple and fast numerical procedure to search for good quantum codes for storing logical qubits in the presence of independent per-qubit noise. In a key departure from past work, we use the worst-case fidelity as the figure of merit for quantifying code performance, a much better indicator of code quality than, say, entanglement fidelity. Yet, our algorithm does not suffer from inefficiencies usually associated with the use of worst-case fidelity. Specifically, using a near-optimal recovery map, we are able to reduce the triple numerical optimization needed for the search to a single optimization over the encoding map. We can further reduce the search space using the Cartan decomposition, focusing our search over the nonlocal degrees of freedom resilient against independent per-qubit noise, while not suffering much in code performance.
\end{abstract}

\maketitle

\section{\label{sec:intro}Introduction}
Interest in building quantum computing devices has grown steadily, with rapid progress in the last few years with the fresh injection of industry support. Current quantum computing devices, like those from IBM, Google, and Rigetti, comprise only a few (at best, tens of) qubits, and are quite noisy. We are right now in the ``NISQ era" \cite{preskill2018quantum}, a term referring to the near-to-intermediate-term situation where physical devices are too noisy and too small to implement regular quantum error correction (QEC) and fault tolerance schemes to deal with the noise in the device. While experimenters work hard to improve the physical qubits and their operation, from the theory side, there is a strong need to find better QEC and fault tolerance schemes with lower resources overheads, essential for the eventual implementation of robust and scalable quantum computing devices. 

The goal of QEC is to offer protection against the loss of information due to noise, e.g., unwanted evolution due to the interaction with the physical environment. Generally, QEC~\cite{shor,  steane,  calderblank,  gottesman, knill} (see also a recent review Ref.~\cite{terhal2015}) tries to store the information to be protected in a special part of the quantum state space with the property that errors due the noise can be identified and their effects removed through a recovery procedure. In its broadest sense, QEC includes passive methods like decoherence-free subspaces and noiseless subsystems~\cite{lidar, shabani,zanardi,holbrook}, where the information is stored and protected in a part of the state space unaffected by the noise and hence requires no recovery operation; more commonly, QEC refers to the situation where active recovery is needed, the typical situation.
 
Much of the existing work on QEC centers around codes capable of removing the effects of \emph{arbitrary} errors on individual qubits, powerful enough to deal with general, even unknown, noise. {The stabilizer codes~\cite{nielsen}, including the well-known Steane code~\cite{steane} and Shor code~\cite{shor}, fall in this category of codes which can correct for single-qubit errors.} 

This generality, however, comes at a price: For example, one cannot find codes capable of correcting an arbitrary error on any single qubit unless one uses at least five physical qubits to encode a one qubit of information \cite{laflamme}. In the current NISQ era, where the question of how to deal with noise is of primary concern, one expects a reasonable level of characterization of the noise afflicting the qubits. In this case, \emph{channel-adapted codes} \cite{FletcherThesis}--- codes tailor-made to deal with the specific noise channel encountered in the physical device--- become of interest. Such codes can be expected, and are known (see, for example, the $4$-qubit code for amplitude-damping noise discovered in Ref.~\cite{leung}), to be less demanding in resources.

The most general formulation of channel-adapted codes requires full knowledge of the noise acting on the entire physical system through the use of process tomography. Process tomography, however, is infamously expensive to perform, with a resource scaling that grows as the fourth power in the dimension of the system, not to mention the difficulties having to do with state-preparation and measurement errors. An accurate joint characterization of multi-qubit noise is hence likely prohibitive in the near future.
Instead, one can make the experimentally well-motivated assumption of independent noise, and regard the full $n$-qubit noise channel as a \emph{tensor product} of $n$ single-qubit channels, hence reducing the problem to characterizing the individual qubit noise, a much simpler task whose difficulty grows only linearly with the number of qubits.

This tensor-product structure, often a good approximation in current devices (see, for example, \cite{tannu}), endows the noise with a local, qubit-by-qubit, character. One then expects to find good code spaces--- regions of the state space resilient to the noise--- among states with a nonlocal structure. This intuition motivated the original idea of quantum codes, storing the information to be protected in states entangled across multiple physical qubits. In our work, we again put it to good use: We focus our search for good channel-adapted codes on states with a nonlocal nature, identified using a Cartan decomposition of the encoding operation.

The question of finding channel-adapted codes can be formulated as an optimization problem~\cite{yamam, Reimpell, fletcher,lidarkosut,wang}, one of finding the combination of code and recovery that optimizes a chosen figure of merit for the given noise channel. When the figure of merit is the \emph{average} entanglement fidelity~\cite{schum}, one only has a double optimization over encoding (of a given block length) and recovery. This problem is known to be tractable via convex optimization techniques~\cite{Reimpell, fletcher, lidarkosut}. However, if the \emph{worst-case fidelity}~\cite{nielsen}---a better figure of merit that assures a minimum performance---is used instead, an additional numerical minimization of the fidelity measure is needed, as one cannot generally write down a closed-form expression for the worst-case fidelity. This leads to a triple optimization problem.

In our work, we focus on finding channel-adapted codes that minimizes the worst-case fidelity for the storage of a single qubit of information, the standard approach to quantum computing. We argue that we can, in practice, reduce the original triple optimization to a single optimization. Specifically, we remove the need for an optimization over the recovery. Instead, we make use of the Petz recovery ~\cite{Petz}, shown to be near-optimal in Ref.~\cite{hui_prabha}. The use of the Petz recovery further permits the use of an analytical expression for the worst-case fidelity for codes encoding a single logical qubit. This removes the need to numerically minimize the fidelity over the state space. Thus, we only need optimize over the encoding operation. Furthermore, a key aspect of this work, we can reduce the difficulty of this remaining numerical optimization over all possible encodings by employing a Cartan decomposition of the encoding operation, motivated by the noise-locality--code-nonlocality dichotomy described earlier. We vary only over the nonlocal pieces of the decomposition, thereby reducing the dimension of the search space. Altogether, these steps give a fast and easy algorithm for finding good channel-adapted codes for the worst-case fidelity, the preferred figure of merit for quantum computing tasks.

{Our paper is organized as follows. In Sec.~\ref{sec:prelim}, we first review the mathematical formalism of channel-adapted codes, and the present the steps that reduce the problem to a single optimization. We describe the the use of the Cartan decomposition to help simplify the search in Sec.~\ref{sec:search}. We present several examples in Sec.~\ref{sec:results} to illustrate our method, and conclude in Sec.~\ref{sec:conc}.}

\section{The optimization problem}\label{sec:prelim}
We begin by describing the various ingredients in our approach to the problem of finding channel-adapted codes for the worst-case fidelity measure. We explain how to reduce the problem to a single optimization, over the encoding operation.

\subsection{Basic formulation}
Consider a physical quantum information processing system of dimension $d$, with Hilbert space $\cH$. The noise acting on the system can be described by a \emph{quantum channel}, i.e., a completely positive (CP) and trace-preserving (TP) linear map, denoted by $\cE$. $\cE$ acts on $\cB(\cH)$, the set of linear operators on $\cH$, $\cE:\cB(\cH)\rightarrow \cB(\cH)$. Its action can be written as $\cE(\,\cdot\,)=\sum_{i=1}^NE_i(\,\cdot\,)E_i^\dagger$, for a set of (non-unique) Kraus operators $\{E_i\}_{i=1}^N$, a structure that assures the CP nature of the map. The Kraus operators further satisfy $\sum_{i=1}^NE_i^\dagger E_i=\id$, for the TP property.

To protect the quantum information from damage by the noise, QEC proposes to store the information---assumed to be a $d_0$-dimensional Hilbert space $\cH_0$ of states---in a $d_0$($\leq d$)-dimensional subspace $\cC$ of $\cH$, the Hilbert space of the physical system. We refer to $\cC$ as the \emph{codespace}. The encoding operation $\cW$---a unitary operation, and hence invertible---is a one-to-one mapping of states from $\cH_0$ to $\cC$, $\cW:\cB(\cH_0)\rightarrow \cB(\cC)\subseteq\cB(\cH)$.  The action of the noise $\cE$ on the encoded state, the output of $\cW$, can then  be regarded as $\cE: \cB(\cC) \rightarrow \cB(\cH)$. After the action of the noise, the QEC protocol applies a suitable recovery map $\cR$, a CPTP map $\cR: \cB(\cH) \rightarrow \cB(\cC)$ that restores the state into the codespace, and in the process removing (hopefully most of) the errors due to the noise. If we want, we can then decode the physical state back into the quantum informational state of $\cH_0$ by applying the decoding operation, $\cW^{-1}$.

Traditionally, the QEC protocol, specified by the pair $(\cW,\cR)$ for given $\cH_0$ and $\cH$, is chosen to satisfy (at least approximately) what are known as the QEC conditions \cite{knill,hui_prabha}, for successful removal of the errors caused by the noise. Here, it is more straightforward to think directly in terms of an optimization problem. For that, we first quantify the performance of a code $\cC$ (or, equivalently, $\cW$) with recovery $\cR$ for the noise process $\cE$ by a measure that compares the output state of the QEC protocol $(\cR\circ\cE)(\rho)$ to the input state $\rho\in\cB(\cC)$. An often-used measure is the fidelity (see, for example, \cite{nielsen}), defined for two states $\rho$ and $\sigma$ as $F(\rho,\sigma) = \tr\sqrt{(\rho^{1/2} \sigma \rho^{1/2})}$. For a pure $\rho$= $|\psi\rangle\langle \psi|$, and for $\sigma\equiv \cM(\rho)$, where $\cM$ is a CPTP map, we write the \emph{square} of the fidelity---a more convenient quantity---as 
\begin{equation}
F^2(|\psi\rangle,\cM) = \langle \psi|\,\cM( |\psi\rangle\langle\psi|) \,|\psi\rangle.
\end{equation}
To characterize the performance of a given pair $(\cW,\cR)$ for noise $\cE$, we use the {\it worst-case fidelity},
\begin{equation}
F^2_{\min}(\cW, \cR;\cE) \equiv \min _{|\psi\rangle \in \cH_{0}} F^{2}(|\psi\rangle,\cW^{-1}\circ \cR \circ \cE \circ \cW). \label{eq:worstcase_fid}
\end{equation}
Above, we have used the fact that the fidelity function $F$ is jointly concave in its arguments, so that the minimum fidelity over all states is always attained on a pure state, i.e., it suffices to minimize over state vectors in $\cH_{0}$. The minimization over $|\psi\rangle\in\cH_0$ usually has to be done numerically unless one has special properties in the problem (as we will see below). Alternatively, one can make use of the \emph{fidelity loss} quantity, 
\begin{equation}
\eta(\cW,\cR;\cE)\equiv 1-F_{\min}^2(\cW,\cR;\cE).
\end{equation}

We can now state the basic formulation of the optimization problem for channel-adapted codes: For given noise $\cE$, and the available dimension $d$ of the physical system, the best code is given by the solution to the following optimization over encoding operation $\cW$ and recovery $\cR$,
\begin{align}\label{eq:optimize}
&\quad~\underset{\cW}{\textrm{argmax}}\,\underset{\cR}{\textrm{argmax}} ~F^2_{\min}(\cW,\cR;\cE)\\
&=\underset{\cW}{\textrm{argmin}}\,\underset{\cR}{\textrm{argmin}} ~\eta(\cW,\cR;\cE)\nonumber\\
&=\underset{\cW}{\textrm{argmax}}\,\underset{\cR}{\textrm{argmax}} \min_{|\psi\rangle\in \cH_0} F^2(|\psi\rangle,\cW^{-1}\circ \cR \circ \cE \circ \cW).\nonumber
\end{align}
This is the triple optimization, over the encoding $\cW$, the recovery $\cR$, and the input state $|\psi\rangle$, mentioned in the introduction.

We note that, in principle, one could also add an optimization over the dimension $d$ of the physical state space used to encode $\cH_0$. For the current situation of independent noise on the physical system, one expects better fidelity with a larger number of physical qubits, as this will gives better ``de-localization" of the information. However, in the current NISQ era, the number of physical qubits available for encoding the information will largely come from practical constraints. We thus take $d$ to be fixed, and find the best $(\cW,\cR)$ for that given $d$.

\subsection{The Petz recovery}
We first reduce this triple optimization problem to a double optimization over $\cW$ and $|\psi\rangle$ only, by choosing a suitable recovery $\cR$. For any noise channel $\cE$ and codespace $\cC$, we choose the corresponding Petz recovery $\cR_{P}$, defined as~\cite{Petz, hui_prabha},
\begin{equation}\label{eq:Petzmap}
\cR_{P}(\cdot) \equiv \sum_{i=1}^{N} PE_{i}^{\dagger} \cE(P)^{-1/2}(\cdot) \cE(P)^{-1/2}E_{i}P, 
\end{equation}
where $\{R_{i}\equiv P E_{i}^{\dagger}\cE(P)^{-1/2}\}_{i=1}^{N}$ constitute the Kraus operators of $\cR_{P}$. Here, $P$ is the projector onto the code space $\cC$, and the inverse of $\cE(P)$ is taken on its support.
The Petz recovery, even though it is usually not the recovery that achieves the smallest worst-case fidelity for given $\cE$ and $\cC$, was shown to be near optimal in \cite{hui_prabha}. Specifically, this near-optimality is captured by the bounds (Corollary 4 of \cite{hui_prabha}),
\begin{equation}
\eta_{\cR_\mathrm{op}}\leq \eta_P\leq \eta_{\cR_\mathrm{op}}{\left[(d_0+1)+O(\eta_{\cR_\mathrm{op}})\right]},
\end{equation}
where $\eta_{\cR_\mathrm{op}}$ and $\eta_P$ are the fidelity losses if we had used the optimal and Petz recoveries, respectively (for the same $\cW$ and given $\cE$). For $d_0=2$, the case we will focus on, we see that the Petz recovery gives a good indicator of the performance of the chosen codespace under the optimal recovery; the use of the Petz recovery does not give a significant deterioration in performance. This justifies our use of the Petz recovery{, a map with a simple analytical form, }as a good proxy for the optimal recovery{, the latter usually accessible only numerically}.

Having fixed the recovery map as $\cR_{P}$, our optimization problem reduces to a double optimization,
\begin{equation}\label{eq:minfidelity}
  \underset{\cW}{\textrm{argmax}} \ \underset{|\psi\rangle\in \cH_{0}}{\text{min}} F^{2}(|\psi\rangle,\cW^{-1}\circ \cR_{P} \circ \cE \circ \cW).
\end{equation}
We denote the fidelity loss for an encoding $\cW$ as $\eta_\cW\equiv \eta(\cW,\cR_P;\cE)$; the optimal encoding $\cW_{\mathrm{op}}$ is then the one that attains $\eta_{\mathrm{op}}\equiv \min_\cW\eta_\cW$.
In this work we search for codes which preserve a qubit worth information, in which case

\subsection{ Fidelity loss for qubit codes}

The optimization problem of Eq.~\eqref{eq:minfidelity} can be further simplified by noting that the worst-case fidelity $\min_{|\psi\rangle\in \cH_0} F^{2}(|\psi\rangle,\cW^{-1}\circ \cR_{P} \circ \cE \circ \cW)$ for encoding $\cW$, or equivalently, the fidelity loss function $\eta_\cW$, has a simple form for the case of qubit codes (i.e., $d_0=2$) with the Petz recovery. Specifically, $\eta_{\cW}$ can be easily computed via eigenanalysis~\cite{hui_prabha}. We recall the steps here, for completeness.

We encode a qubit $\cH_0$ into a two-dimensional codespace $\cC$. For an orthonomal basis $\{|v_1\rangle,|v_2\rangle\}$ on $\cC$, the Pauli basis $\{\sigma_\alpha\}_{\alpha=0,x,y,z}$ (orthogonal but not normalized) for operators on $\cC$ can be defined in the usual way as
\begin{align}\label{eq:Pauli}
\sigma_0 =& |v_1\rangle\langle v_1|+|v_2\rangle\langle v_2|=P \equiv \id_2,\\
\sigma_x =& |v_1\rangle\langle v_2|+|v_2\rangle\langle v_1|, \nonumber \\
\sigma_y =& -i(|v_1\rangle\langle v_2|-|v_2\rangle\langle v_1|), \nonumber\\
\textrm{and}\quad\sigma_z =& |v_1\rangle\langle v_1|-|v_2\rangle\langle v_2|.
\end{align}
Codestates $\rho\in\cB(\cC)$ can then be described using the Bloch-ball representation,
\begin{equation}
\rho = \dfrac{1}{2}(\id_2 + \mathbf{s} . \bm{\sigma}),
\end{equation} 
where $\mathbf{s}=(s_x,s_y,s_z)$ is a real $3$-dimensional vector---the Bloch vector for $\rho$---with Euclidean length $|\mathbf{s}|\leq 1$, and $\bm{\sigma}=(\sigma_x,\sigma_y,\sigma_z)$.

Consider the channel $\cM:\cB(\cC)\rightarrow \cB(\cC)$ composed from the noise followed by the Petz recovery, acting solely on the codespace, $\cM\equiv\cR_P\circ\cE\circ\cP$. $\cP(\cdot)\equiv P(\cdot)P$ is the map that enforces the pre-condition that we start in the codespace. $\cM$ is both trace-preserving [$\cM^\dagger (P)=P$] and unital [$\cM(P)=P$], and its action can be expressed, in the Pauli operator basis, as the matrix
 \begin{equation}\label{eq:tmatrix}
M  = {\left(\begin{array}{c|c}
1 & 0~~0~~0\\
\hline
0 &  \\
0 & T \\
0 & 
\end{array}\right)},
\end{equation}
with real matrix entries $M_{\alpha\beta}\equiv \frac{1}{2}\tr\{\sigma_\alpha\cM(\sigma_\beta)\}$; $T$ is a $3\times 3$ matrix of the $\alpha,\beta=x,y,z$ entries. The action of $\cM$ on an input state $\rho\in\cB(\cC)$ can then be expressed in terms of the action on the Bloch vector as $\mathbf{s}\mapsto \mathbf{s}'\equiv T\mathbf{s}$.
The fidelity loss $\eta_\cW$ (for given encoding $\cW$ that defines the $\cC$ subspace) is then, by straightforward algebra,
\begin{equation}
\eta_\cW=\max_{\mathbf{s},|\mathbf{s}|=1}\tfrac{1}{2}(1-\mathbf{s}^\mathrm{T}T_\mathrm{sym}\mathbf{s})=\tfrac{1}{2}{\left[1-t_{\min}(\cW)\right]},
\end{equation}
where $T_\mathrm{sym}\equiv \frac{1}{2}(T+T^\mathrm{T})$, and $t_{\min}(\cW)$ is the smallest eigenvalue of $T_\mathrm{sym}$. Here, the superscript $\mathrm{T}$ denotes the transpose operation.

\subsection{A single numerical optimization}
In this way, we have reduced the minimization needed to compute the worst-case fidelity in Eq.~\eqref{eq:minfidelity} to a simple diagonalization of a $3\times 3$ matrix and taking the smallest eigenvalue. We thus finally have only a single optimization left to do to find the best channel-adapted code,
\begin{equation}
  \underset{\cW}{\textrm{argmin}}~\frac{1}{2}{\left[1-t_{\min}(\cW)\right]}.
\end{equation}

The optimization over the encoding $\cW$ has to be done numerically. We parameterize the search space as follows. Every codespace $\cC$ is specified by $d_0$ orthogonal pure states in $\cH$, forming a basis for $\cC$. Varying over the codespace can then be thought of as starting with a fixed basis with $d_0$ elements, and then applying a rotation of the basis, via a unitary operator $U$, in the full $d$-dimensional Hilbert space of the physical system. Choosing different $\cC$'s then corresponds to choosing different unitary operators $U$. The search space is then the set of all $d$-dimensional unitary operators, a space specified by $d^2$ real parameters.

Observe that $t_{\min}(\cW)$ has to be computed numerically for each $\cW$. This means that we do not have a closed-form expression for the gradient of our objective function, so that standard optimization methods that require a formula for the gradient do not work. This is easily solved, however, by going to methods that estimate the gradient numerically within the gradient-descent algorithm. A well-known approach, the one that we used here, is the Nelder-Mead search technique (also known as the downhill simplex method; see, for example, {Refs.~\cite{nelderpaper, NumericalRecipes}}).

\section{Simplifying the search: The Cartan decomposition}\label{sec:search}

As stated earlier, the optimization over $\cW$ involves a $d^2$-dimensional search. For $n$-qubit physical systems, the typical experimental scenario, where $d=2^n$, the search space dimension grows exponentially with $n$. It would hence be useful to further reduce the complexity of the search by considering a restricted search over the set of encoding unitaries $\cW$. For that, we recall our focus, as motivated in the introduction, on noise channels with a tensor-product structure over the $n$ qubits. This local structure in the noise suggests  the use of codes with a nonlocal nature. To separate the nonlocal pieces of the unitary search space from the local pieces, we make use of the Cartan decomposition, as described in Sec.~\ref{sec:cartan}.

Our search space, originally comprising elememts in the unitary group $U(2^{n})$ for an $n$-qubit code, can be restricted to elements of the special unitary group $SU(2^{n})$ without loss of generality. We then use the Cartan decomposition originally proposed in~\cite{khaneja_glaser}, whereby any $n$-qubit unitary is realised as a product of single-qubit (local) and multi-qubit (nonlocal) unitaries. The specific paramterization we use is due to~\cite{cartan}, where the standard Pauli basis is employed to decompose an arbitrary element of  $SU(2^{n})$ in terms of its local and nonlocal parts in an iterative fashion.

\subsection*{Cartan form of the encoding unitary}\label{sec:cartan}

Recall that the special unitary group $SU(d)$---the group of $d \times d$ complex matrices with determinant one---forms a real Lie group of dimension $d^{2}-1$. Let $\mathfrak{SU}(d)$ denote the corresponding Lie algebra, the algebra of traceless anti-Hermitian $d\times d$ complex matrices with the Lie bracket $-\upi[\,\cdot\,,\,\cdot\,]$, i.e., $(-\upi)$ times the commutator.

The central idea behind the Cartan form is the fact that any element of $SU(2^{m})$ can be represented, up to local unitaries, using elements of two Abelian subalgebras $\mathfrak{h_{m}}$ and $\mathfrak{f_{m}}$ ($m=2, 3, \ldots , n$) of $\mathfrak{SU}(2^{m})$. This was shown in~\cite{khaneja_glaser} via an iterative decomposition of the form $U = U' H U''$, where $H$ is generated alternately from elements of $\mathfrak{h_{m}}$ and $\mathfrak{f_{m}}$, while $U'$ and $U''$ belong to the subgroup of $SU(2^{m})$ generated by a subalgebra orthogonal to $\mathfrak{h_{m}}$ and $\mathfrak{f_{m}}$. The exact structure of the decomposition depends on the choice of an appropriate basis for $\mathfrak{SU}(2^{m})$ that can be obtained recursively for $m=2,3,\ldots,n$. 

For example, for $n=2$, one can use twofold tensor products of the single-qubit Pauli operators ($I,X,Y,Z$) as basis elements for $\mathfrak{SU}(4)$. Using this basis to partition $\mathfrak{SU}(4)$ into orthogonal subspaces leads to an identification of the Abelian subalgebras $\mathfrak{h}_{2} = {\rm span}\{XX,YY, ZZ\}$ and $\mathfrak{f}_{2} = \{0\}$. This leads to the well-known Cartan form for $U \in SU(4)$~\cite{Zhang2003, Rezakhani2004},
\begin{equation}
 U = (U_{1}\otimes U_{2})\,\upe^{-\upi  (c_{1}XX + c_{2}YY + c_{3}ZZ)} (U_{3}\otimes U_{4}), \label{eq:su2}
 \end{equation}
where $U_{1}, U_{2}, U_{3}$, and $U_{4} \in SU(2)$ are local, single-qubit unitaries, and $c_{1}, c_{2}$, and $c_{3}$ are scalar parameters for the nonlocal operators.
 
Following this intuition from $SU(2)$, Ref.~\cite{cartan} showed that a basis comprising $n$-fold tensor products of the single-qubit Pauli basis can be obtained for any $\mathfrak{SU}(2^{n})$ by an iterative process, which partitions $\mathfrak{SU}(2^{n})$ into Abelian subalgebras $\mathfrak{h}_{n}$ and $\mathfrak{f}_{n}$. The Cartan decomposition of any $n$-qubit unitary operator can then be obtained as follows: \\[1ex]
{\textbf{Cartan decomposition}~\cite{cartan}\textbf{.~}}\textit{Any $U \in SU(2^{n})$, for $n > 2$ can be decomposed as,
\begin{equation}\label{eq:cartan}
G = K^{(1)} F^{(1)} K^{(2)} J K^{(3)} F^{(2)} K^{(4)}.
\end{equation}
Here, each $K^{(i)}$ denotes a product operator from $SU(2^{n-1}) \otimes SU(2)$, $F^{(j)}\equiv\exp(-\upi f^{(j)})$ and $J\equiv \exp{(-\upi h)}$ are unitary operators \emph{nonlocal} on the entire $n$-qubit space, with $h \in \mathfrak{h}_{n}$ and $f^{(j)} \in \mathfrak{f}_{n}$. The decomposition can be applied recursively, to further decompose each $SU(2^m)$ operator in $K^{(i)}$, in the form of Eq.~\eqref{eq:cartan}, for $m=2,3,\ldots,n-1$.}\\

As in the $n=2$ case, the Cartan decomposition for $n >2$ again separates out the local and nonlocal degrees of freedom  in an iterative fashion. However, for $n > 2$, a second Cartan decomposition is required in order to identify the factors that are nonlocal on the entire $n$-qubit space. This stems from the fact that a pair of nontrivial Abelian subalgebras $\mathfrak{h}_{n}, \mathfrak{f}_{n}$ maybe identified for any $n >2$, and this leads to a two{-}step decomposition. First, using the generators of the subalgebra $\mathfrak{h}_{n}$, we obtain the unitary $J\in SU(2^{n})$, as well as $U', U'' \in SU(2^{n})$, such that $G = U' J U''$ for any $G \in SU(2^{n})$, $n>2$. Further Cartan decompositions of $U'$ and $ U''$ using the generators of $\mathfrak{f}_{n}$ gives the form $ G= K^{(1)} F^{(1)} K^{(2)} J K^{(3)} F^{(2)} K^{(4)}$, where the operators $K^{(i)} \in SU(2^{n-1})\otimes SU(2)$ are no longer nonlocal on the entire $n$-qubit space. Starting with the bases for $\mathfrak{h}_{2}, \mathfrak{f}_{2}$ identified above, Ref.~\cite{cartan} provides a simple recursive prescription to identify the bases for the subalgebras $\mathfrak{h}_{n}, \mathfrak{f}_{n}$, for any $n>2$. 

To illustrate how the above prescription can be used to obtain a nice parameterization of the encoding unitaries for QEC, we explicitly write down the Cartan form for $n=3$ and $4$. Any element of $SU(2^{3})$ can be constructed using the formalism in Eq.~\ref{eq:cartan} as,
\begin{align}\label{eq:su8}
U &= K^{(1)} F^{(1)} K^{(2)} J K^{(3)} F^{(2)} K^{(4)},\\
F^{(i)} &= \upe^{-\upi (c^{(i)}_{1}XXZ + c^{(i)}_{2}YYZ + c^{(i)}_{3}ZZZ)}, \nonumber \\ 
\textrm{and }\quad J &= \upe^{-\upi(a_{1}XXX+a_{2}YYX+  a_{3}ZZX+ a_{4}IIX)}.\nonumber
\end{align}
As stated above, $K^{(j)} \in SU(4) \otimes SU(2)$, and each element in $SU(4)$ can be obtained similarly from Eq.~\ref{eq:su2}. Recall that the standard description of any unitary in $SU(2^{3})$ requires $63$ real parameters, whereas the recursive Cartan decomposition described in Eq.~\ref{eq:su8} requires a total of $82$ real parameters. However, the key advantage of using the Cartan parameterization is that the nonlocal factors of any unitary in $SU(2^{3})$ are easily described in terms of $22$ real parameters,namely, the set of ten real parameters $\{a_{1}, a_{2}, a_{3}, a_{4}, c^{(i)}_{1}, c^{(i)}_{2}, c^{(i)}_{3}\}$, along with three real parameters for each of the four $SU(4)$ factors. 

Similarly, we note that any element $ U \in SU(2^{4})$ can be decomposed as,
\begin{align}\label{eq:su16}
 U &= K^{(1)} F^{(1)} K^{(2)} {J} K^{(3)} F^{(2)} K^{(4)}, \nonumber \\ 
 F^{(i)} &=\exp\bigl(-{\upi}(c^{(i)}_{1} XXIZ+c^{(i)}_{2}YYIZ\nonumber\\
 &~\quad\qquad + c^{(i)}_{3} ZZIZ + c^{(i)}_{4} IIXZ + c^{(i)}_{5} XXXZ\nonumber\\
 &~\quad\qquad + c^{(i)}_{6} YYXZ + c^{(i)}_{7} ZZXZ) \bigr),\quad i=1,2, \nonumber \\ 
\textrm{and } J &= \exp\bigl(-{\upi} (a_{1} IIIX+a_{2} XXIX + a_{3} YYIX  \nonumber \\  
 &~\quad\qquad + a_{4} ZZIX+ a_{5} IIXX+ a_{6}XXXX+ \nonumber \\  
 &~\quad\qquad + a_{7} YYXX + a_{8} ZZXX) \bigr), 
\end{align}
where $ K^{(j)}$ $\in$ $SU(8)\otimes SU(2)${, $j=1$, 2, 3, and 4}. Such a decomposition requires a total of $362$ real parameters including the set $\{a_{1}, a_{2}, a_{3},\ldots, a_{8}, c^{(i)}_{1}, c^{(i)}_{2},\ldots c^{(i)}_{7}\}$ which parameterizes the fully nonlocal factors.
 
We can further simplify our numerical search by fixing the local components in the Cartan form and searching only over the nonlocal degrees of freedom. This greatly reduces the dimension of our search space {and allows us to search much more quickly, compared to the unstructured search}. This restriction only to nonlocal degrees of freedom, as we see in our examples in Sec.~\ref{sec:results}, does not lead to substantial loss in fidelity.

As an aside, we note that choosing the local unitaries in the decomposition appropriately allows us to construct encoding unitaries with simple structures that permit easy circuit implementations of the encoding procedure. 
An example of such a \emph{structured} encoding would be to set all the $K^{(i)}$s in Eq.~\ref{eq:cartan} to be the identity operator, thus reducing the form of the encoding unitary to $U =F^{(1)} J F^{(2)}$. This implies the following form for the encoding unitary in the case of $SU(2^3)$,
\begin{equation}
U=
   \begin{bmatrix}
\ast  & \ast&   0 &   0 &   0 &   0 &   \ast & \ast &\\
  \ast & \ast &   0 &   0 &   0 &   0 &  \ast & \ast &\\
   0 &   0 & \ast & \ast & \ast & \ast &   0 &   0 &\\
     0 &   0 &  \ast & \ast & \ast & \ast &   0 &   0 &\\
   0 &   0 & \ast & \ast & \ast &\ast & 0 &   0 &\\
   0 &   0 &  \ast  & \ast & \ast & \ast &   0 &   0 &\\
  \ast & \ast &   0 &   0 &   0 &   0 &  \ast & \ast &\\
   \ast & \ast &   0 &   0 &   0 &   0 &  \ast & \ast &
  \end{bmatrix} , \label{eq:U_8}
  \end{equation}
where $\ast$ refers to some non-zero complex number.  Such a structured encoding $U$ with only non-local Cartan factors is easy to implement using only single{-}qubit gates and the CNOT gate, as explained in Appendix~\ref{sec:circuit}.

\section{Examples}\label{sec:results}

Let us now demonstrate the performance of our approach to finding good codes through a few examples. We consider $n$-qubit noise channels of the form $\cE=\cE_1\otimes\cE_2\otimes\ldots\otimes \cE_n$, where $\cE_i$ is a single-qubit channel on the $i$-th qubit. The first few examples are for the cases where all the $\cE_i$s are the same channel, corresponding to the common experimental situation where all the qubits see the same environment and hence undergo the same noise dynamics. We look at three examples: $\cE_i$ is the amplitude-damping channel, the rotated amplitude-damping channel, and an arbitrary (randomly chosen, no special structure) single-qubit channel. 

In each example, we use the form of the encoding unitary in Eq.~\ref{eq:cartan} to perform both {\it unstructured} as well as {\it structured} search over the space of all encoding unitaries.
In an unstructured search, we retain the general form of the encoding unitary in Eq.~\ref{eq:cartan}, using \emph{all} parameters, local and nonlocal, in our search. For the structured search, we have two different approaches. In the case of a structured search with \emph{trivial} local unitaries, we set all the local ($SU(2)$) unitaries in the Cartan decomposition in Eq.~\ref{eq:cartan} equal to the identity, and search only over the nonlocal parameters in Eq.~\eqref{eq:cartan}. For example, while searching over the $4$-qubit space, we retain the nontrivial $3$-qubit nonlocal pieces as well as the $2$-qubit pieces, but set all the single-qubit unitaries to identity. In addition, for the example of Sec.~\ref{sec:rotated_AD}, we also implement structured search with \emph{nontrivial} local{ ($SU(2)$)} unitaries where the choice of the local unitaries in the Cartan decomposition is guided by the structure of the channel. 

\subsection{Amplitude-damping channel}\label{sec:amp_damp}

\begin{figure}
\includegraphics[trim=8mm 12mm 15mm 5mm, clip, width=\columnwidth]{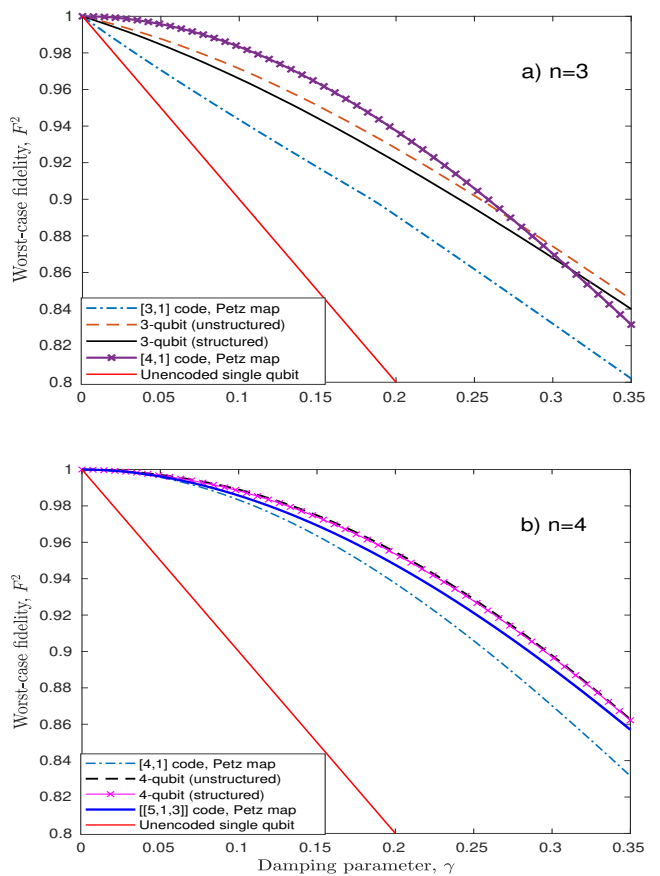}
\caption{\label{fig:3-4qubit}Performance of $n$-qubit codes for amplitude{-}damping, using the Petz recovery [Eq.~\eqref{eq:Petzmap}] with structured and unstructued encodings, for (a) $n=3$, and (b) $n=4$.} 
\end{figure}

The single-qubit amplitude-damping channel $\cE_\mathrm{AD}$ is described by a pair of Kraus operators in the computational ($\sigma_z$) basis $\{|0\rangle,|1\rangle\}$,
{
\begin{align}\label{eq:amplitudedamping}
E_{0}  &= |0\rangle\langle 0|+ \sqrt{1-\gamma}\,|1\rangle\langle 1|,\nonumber\\
\textrm{and}\quad E_1 &= \sqrt{\gamma}\,|0\rangle\langle 1|,
\end{align}
}
where $E_1$ flips the $|1\rangle$ state to the $|0\rangle$ state, imitating a ``decay" to the $|0\rangle$ state; the deviation of $E_0$ from the identity is needed for the trace-preserving nature of the channel. We perform the numerical search outlined in Sec.~\ref{sec:search} and obtain optimal encodings for $\cE=(\cE_\mathrm{AD})^{\otimes n}$, for $\gamma\ll 1$, for $n=3$ and $4$. We compare the performance of the codes we find with various known codes; see Fig.~\ref{fig:3-4qubit}. The $[3,1]$ approximate code~\cite{langshor} is the span of the states,
\begin{equation}\label{eq:3qubit}
 |0_{L}\rangle = \tfrac{1}{\sqrt 2}(|000\rangle +|111\rangle), \quad
 |1_{L}\rangle = \tfrac{1}{\sqrt 2}(|100\rangle +|011\rangle);
 \end{equation}
the $[4,1]$ approximate code~\cite{leung} is the span of
\begin{equation}\label{eq:4qubit}
 |0_{L}\rangle = \tfrac{1}{\sqrt 2}(|0000\rangle +|1111\rangle), \quad
 |1_{L}\rangle = \tfrac{1}{\sqrt 2}(|1100\rangle +|0011\rangle).
 \end{equation}
Fig.~\ref{fig:3-4qubit} shows that the numerically obtained codes via structured (with trivial local unitaries) and unstructured search outperform the {\it approximate} code of the same length in Eq.~\eqref{eq:3qubit} and Eq.~\eqref{eq:4qubit} respectively. We observe that the performance of the $4$-qubit optimal codes is even better than the standard $[[5,1,3]]$ code, as seen in Fig.~\ref{fig:3-4qubit}(b). In both cases we have also plotted the worst-case fidelity for a single unprotected qubit under the noise channel. The fidelity of the unencoded qubit falls off linearly with the noise parameter, thus demonstrating the advantage of using the 3- or 4-qubit  codes found using our procedure 

Appendix~\ref{app:numericalcodes} contain{s} the codewords for the optimal $3$,$4$-qubit codes found in our search. We also provide the encoding circuit corresponding to the optimal, structured $3$-qubit code, as an example of how the codes that emerge out of the structured search admit simple encoding circuits. Finally, we note that our numerical search procedure is indeed fast: The unstructured search for a specific value of damping parameter $\gamma$ takes a few hundred seconds on a standard desktop computer, while each structured search takes only a few milliseconds on the same computer.

\subsection{Rotated amplitude-damping channel}\label{sec:rotated_AD}

\begin{figure}
\includegraphics[trim=5mm 3mm 9.3cm 3mm, clip, width=\columnwidth]{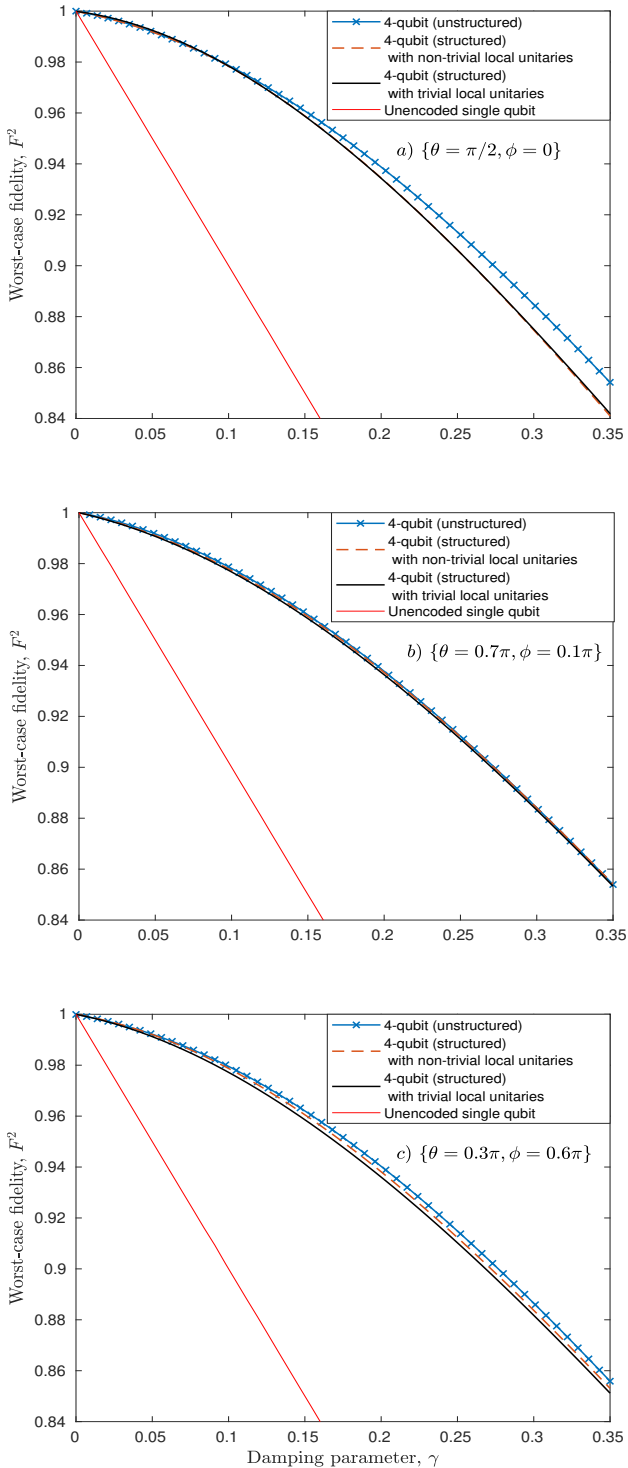}
\caption{\label{fig2}Approximate $4$-qubit codes for damping along different directions in the Bloch sphere: (a) the $x$ direction (spherical coordinates $\{\theta,\phi \}$= $\{\pi/2,0\}$); (b) along the direction $\{\theta,\phi\}=\{0.7\pi,0.1\pi\}$; and (c) along the direction $\{\theta,\phi\}= \{0.3\pi,0.6\pi\}$.}
\end{figure} 

As our Cartan decomposition uses the Pauli basis, it is important to test if our numerical search is robust against noise not aligned along the axes used to define the Pauli basis. For this, we try out amplitude-damping channels where the damping is no longer in the $\sigma_z$ basis. Specifically, we consider the single-qubit \emph{rotated} amplitude-damping channel $\cE_\mathrm{RAD}$ described by the Kraus operators,
\begin{align} \label{eq:arbitrary}
E'_{0} &= |v\rangle\langle v| +\sqrt{1-\gamma} \, \vert v^{\perp}\rangle \langle v^{\perp}\vert, \nonumber \\
\textrm{and }\quad E'_{1} &= \sqrt{\gamma} \, \vert v \rangle \langle v^{\perp} \vert ,
\end{align}
where $\{|v\rangle,|v^\perp\rangle\}$ is a pair of orthonormal vectors on the Bloch sphere. Such a pair of vectors can be parameterized with respect to the $\{|0\rangle, |1\rangle\}$ basis using spherical coordinates,
\begin{eqnarray}\label{eq:parametrise}
|v\rangle &=& \cos(\theta/2) |0\rangle  + \upe^{\upi \phi} \sin (\theta/2) |1\rangle, \nonumber \\
\textrm{and }\quad |v^{\perp}\rangle &=& - \upe^{-\upi \phi} \sin(\theta/2) |0 \rangle + \cos(\theta/2) |1\rangle ,
\end{eqnarray}
with $\theta$ $\in$ $[0,\pi]$, $\phi$ $\in$ $[0,2\pi]$. The values of $\{\theta,\phi\}$ thus determine the damping direction. 

We present numerical search results for the amplitude-damping channel aligned along three different directions in Fig.~\ref{fig2}. In all three examples, the structured search with nontrivial local unitaries was implemented by fixing the local unitaries as $U \equiv (|v\rangle \langle 0|+|v^{\perp}\rangle \langle 1|)$ $\in$ $SU(2)$. For example, when the damping noise is aligned along the $x$-direction on the Bloch sphere, the basis $\{|v\rangle, |v^{\perp}\rangle\}$ is the eigenbasis of $\sigma_{x}$ and the local unitaries are fixed to be the Hadamard gate, which rotates the $\{|0\rangle, |1\rangle\}$ basis to the $\{|+\rangle, |-\rangle\}$ basis. 

Fig.~\ref{fig2}(a) shows the performance of different codes when the damping is with respect to the $\sigma_{x}$ eigenstates, whereas Figs.~\ref{fig2}(b) and (c) present the results for choices of damping direction $|v\rangle$ not aligned with one of the standard Pauli axes. In all three cases, we observe that the codes obtained using the unstructured search offer only slightly better fidelity than the codes obtained using the structured searches. Furthermore, the codes obtained using non-trivial local unitaries are often distinct from, and offer better fidelity compared to the codes obtained using trivial local unitaries in the search. Once again, our search procedure is efficient, with the structured and unstructured searches taking between tens to hundreds of seconds on a standard desktop computer. As in the earlier case, we have also compared the performance of the $4$-qubit codes with the fidelity of the single unprotected qubit.

\subsection{Random local noise}
As a third example of the usefulness of our numerical search procedure, we search for good codes for $\cE^{\otimes n}$ where $\cE$ is a randomly chosen single-qubit channel. A random qubit channel $\Phi$ is generated from a Haar-random unitary on the qubit and a single-qubit ancilla initialized to the state $|0\rangle$; the unitary acts jointly on the qubit and the ancilla, after which the ancilla is traced out, giving a single-qubit channel. We then admix $\Phi$ with the identity channel to give a family of qubit noise channels $\cE$, for different $\alpha\in[0,1]$,
\begin{equation}\label{eq:weaknoise}
\cE(\cdot)=(1-\alpha)(\cdot) + \alpha \Phi(\cdot).
\end{equation}
$\alpha$ parametrizes the noise strength; for small values of $\alpha$, $\cE$ describes weak noise, the practically relevant case.

\begin{figure}
\includegraphics[trim=5mm 9cm 12mm 5mm, clip, height=7cm, width=\columnwidth]{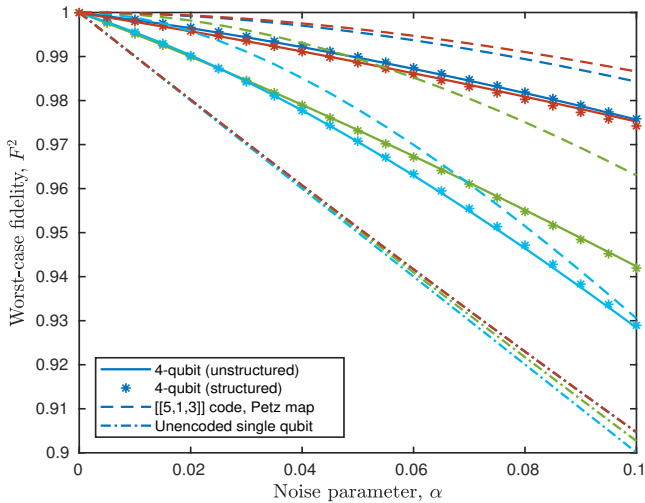}
\caption{Performance of $4$-qubit numerical code and the $[[5,1,3]]$ code for random local noise.}
\label{fig:random}
\end{figure}

Using our numerical search procedure we now obtain optimal $4$-qubit codes for the class of random local noise channels described by Eq.~\eqref{eq:weaknoise} in the weak noise regime, $\alpha \in [0,0.1]$. For each choice of random $\Phi$, and hence $\cE$ for varying $\alpha$, we use our numerical AQEC approach to identify good $4$-qubit codes for the 4-qubit channel $\cE^{\otimes 4}$.

Fig.~\ref{fig:random} shows the fidelities obtained for the optimal codes for four random choices of $\Phi$. We compare the performance of the four-qubit codes obtained via structured and unstructured searches with the performance of the $[[5,1,3]]$ code, for each random channels. The recovery procedure used in each case is the corresponding Petz recovery. 

In Fig.~\ref{fig:random}, we observe that the best $4$-qubit codes---structured or unstructured---have fidelities linear in the noise strength $\alpha$, suggesting that the $4$-qubit Hilbert space might not be sufficient to distinguish among the no-error case and the eight single-qubit errors arising from the weak noise $\cE$. This issue is clearly resolved when we use the $[[5,1,3]]$ code and the corresponding Petz recovery, since we the fidelity is now quadratic to leading order in $\alpha$. Finally, we note that the $4$-qubit codes do yield a better worst-case fidelity than the single unencoded qubit under the action of the noise channel.

\bigskip
\bigskip
\section{Conclusions}\label{sec:conc}
We described a numerical search algorithm to find good quantum codes, using the worst-case fidelity as the figure of merit. By choosing the recovery map as the Petz recovery, we reduced the general problem from a triple optimization over the encoding, recovery, and input states, to a single optimization over the encoding map only. Furthermore, the use of the Cartan decomposition, motivated by the typical scenario of independent per-qubit noise, allowed for a reduction of the search space to structured encodings, with performance comparable with the more expensive unstructured ones, as illustrated by our examples.           

{In our search for the quantum codes we have assumed that the encoding/decoding gates are perfect. This is of course the standard assumption in any discussion on QEC codes in the literature. However, the gates used to do the error correction are indeed the same ones as those used to do computation, so this assumption has to be relaxed, taking us into the domain of fault-tolerant quantum computing~\cite{gottesman_FT}. Our work deals with the first step of finding the optimal code for a given noise process. Extending this framework to fault tolerance is the next step for future work. In this context, our work provides an easy platform to explore and discover a large number of good candidate codes, which can then be individually examined to find the one that can be most easily implemented in a fault-tolerant manner.}

{Furthermore, the ability to identify channel-adapted codes that involve fewer qubits than the stabilizer codes targeting arbitrary noise, might suggest that the corresponding encoding/decoding circuits might also be smaller in size. In fact, the well-known $4$-qubit channel-adapted code due to Leung et al~\cite{leung} does have a much simpler encoding circuit~\cite{FletcherThesis} than the $5$-qubit stabilizer code, and the encoding circuit is made up of only Clifford gates.}

It would be interesting to study how our procedure extends to the case of passive error suppression techniques such as decoherence-free subspaces (DFS)and noiseless subsystems. We note here that it is indeed straightforward to extend our search procedure to check for the existence of DFS for a given noise model. For example, motivated by existence of a DFS for correlated amplitude-damping noise~\cite{correlated_AD} on two qubits, we examined the case of the $3$-qubit and $4$-qubit correlated amplitude-damping noise channels. Preliminary results suggest that our unstructured search procedure, where we make use of a full, unstructured parameterization of the encoding unitary, may work well and can identify the subspaces corresponding to a DFS for both cases. 

%
\def\germ{\frak} \def\scr{\cal} \ifx\documentclass\undefinedcs
  \def\bf{\fam\bffam\tenbf}\def\rm{\fam0\tenrm}\fi 
  \def\defaultdefine#1#2{\expandafter\ifx\csname#1\endcsname\relax
  \expandafter\def\csname#1\endcsname{#2}\fi} \defaultdefine{Bbb}{\bf}
  \defaultdefine{frak}{\bf} \defaultdefine{=}{\B} 
  \defaultdefine{mathfrak}{\frak} \defaultdefine{mathbb}{\bf}
  \defaultdefine{mathcal}{\cal}
  \defaultdefine{beth}{BETH}\defaultdefine{cal}{\bf} \def\bbfI{{\Bbb I}}
  \def\mbox{\hbox} \def\text{\hbox} \def\om{\omega} \def\Cal#1{{\bf #1}}
  \def\pcf{pcf} \defaultdefine{cf}{cf} \defaultdefine{reals}{{\Bbb R}}
  \defaultdefine{real}{{\Bbb R}} \def\restriction{{|}} \def\club{CLUB}
  \def\w{\omega} \def\exist{\exists} \def\se{{\germ se}} \def\bb{{\bf b}}
  \def\equivalence{\equiv} \let\lt< \let\gt>

\appendix

\section{Structured Encodings and Encoding Circuits}\label{sec:circuit}
{Here, we} describe simple circuits by means of which a structured encoding of the form given in Eq.~\eqref{eq:U_8} can be implemented. Recall that the Cartan decomposition for such a unitary is given by,
\begin{equation}\label{eq:nonlocalunitary}
U=F^{(1)}J F^{(2)},
\end{equation}
where the unitary operators $\{F^{(1)},{J},F^{(2)}\}$ are constructed from elements of an Abelian subgroup of the $n$-fold Pauli group. For the case of $SU(2^{3})$, these operators are given by{ [see Eq.~\eqref{eq:su8}]},
\begin{eqnarray}
 F^{(1)} &=& \upe^{-\upi(c_{1}XXZ+c_{2}YYZ+c_{3}ZZZ)} \nonumber \\ 
  &=& \upe^{-\upi c_{1}XXZ} \upe^{-\upi c_{2}YYZ} \upe^{-\upi c_{3}ZZZ} \nonumber \\ 
F^{(2)}&=& \upe^{-\upi c_{4}XXZ} \upe^{-\upi c_{5}YYZ} \upe^{-\upi c_{6}ZZZ} \nonumber \\ 
J &=& \upe^{-\upi(a_{1}XXX+a_{2}YYX+a_{3}ZZX+a_{4}IIX)} \nonumber \\
 &=&\upe^{-\upi a_{1}XXX} \upe^{-\upi a_{2}YYX} \upe^{-\upi a_{3}ZZX}\upe^{-\upi a_{4}IIX} . \label{eq:abeliangroup2}
\end{eqnarray}

Following a simple prescription in~\cite{swaddle}, we can construct quantum circuits to implement $F^{(1)}, F^{(2)}$  and $J$ by suitably combining the simple circuits given in Fig.~\ref{fig:circuits}. The overall encoding uintary $U$ given in Eq.~\eqref{eq:nonlocalunitary} is then composed from the circuits for $F^{(1)}$, $F^{(2)}$ and $J$ as given in Fig.~\ref{fig:circuits2}.

\begin{figure}[h!]
\flushleft
(a) \hspace*{-0.7cm}\Qcircuit @C=1em @R=.7em {
&&&&& \qw & \ctrl{1} & \qw & \qw  &\qw & \qw  & \ctrl{1}    &\qw \\
&&&&& \qw & \targ  &\qw  & \ctrl{1} & \qw & \ctrl{1} & \targ  &\qw \\
&&&&& \qw &\qw  &\qw &\targ  &\gate{R_{z}(-\alpha_{1})}  &\targ &\qw  &\qw 
}
\vspace*{0.5cm}

(b) \hspace*{-0.4cm}\Qcircuit @C=1em @R=.7em {
&&&& \gate{H} & \ctrl{1} & \qw & \qw  &\qw & \qw  & \ctrl{1}   & \gate{H^{\dagger}} &\qw \\
&&&& \gate{H} & \targ  &\qw  & \ctrl{1} & \qw & \ctrl{1} & \targ & \gate{H^{\dagger}} &\qw \\
&&&& \qw &\qw  &\qw &\targ  &\gate{R_z(-\alpha_{1})}  &\targ &\qw & \qw &\qw
}
\vspace*{0.5cm}

(c) \hspace*{-0.4cm}\Qcircuit @C=1em @R=.7em {
&&&& \gate{S} & \ctrl{1} & \qw & \qw  &\qw & \qw  & \ctrl{1}   & \gate{S^{\dagger}} &\qw \\
&&&& \gate{S} & \targ  &\qw  & \ctrl{1} & \qw & \ctrl{1} & \targ & \gate{S^{\dagger}} &\qw \\
&&&& \qw &\qw  &\qw &\targ  &\gate{R_z(\alpha_{1})}  &\targ &\qw & \qw &\qw
}
\label{fig:circuits}
\caption{ Quantum circuit implementing {(for $\alpha_{1} \in \mathbb{R}$) (a) the gate $e^{(-i \alpha_{1} Z \otimes Z \otimes Z)}$; (b) the gate $e^{(-i \alpha_{1} X \otimes X \otimes Z)}$; and (c) the gate $e^{(-i \alpha_{1} Y \otimes Y \otimes Z)}$. Here, $H\equiv|+\rangle\langle 0|+|-\rangle\langle 1|$ is the Hadamard gate, and $S\equiv \frac{1}{\sqrt 2}(\id+\upi \sigma_x)$.}}
\end{figure}
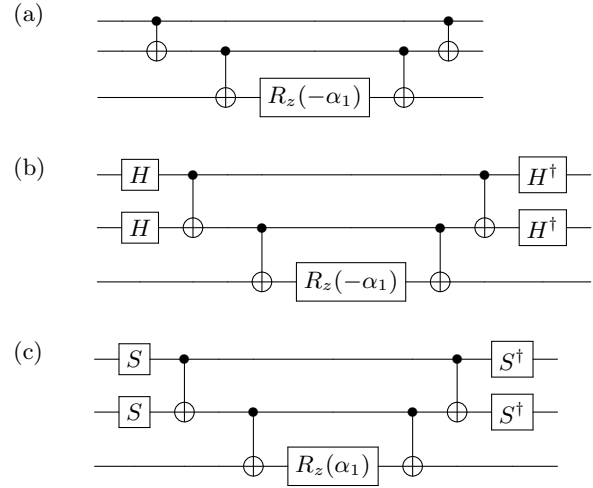

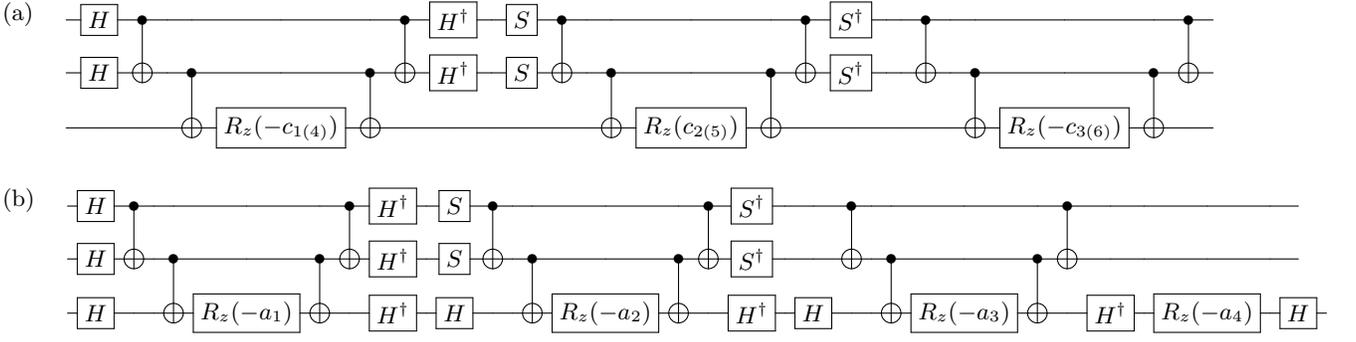
\begin{figure*}
\flushleft
(a) \quad \Qcircuit @C=0.6em @R=0.7em {
& \gate{H} & \ctrl{1} & \qw & \qw  &\qw & \qw  & \ctrl{1}   & \gate{H^{\dagger}} &\qw & \gate{S} & \ctrl{1} & \qw & \qw  &\qw & \qw  & \ctrl{1}  & \gate{S^{\dagger}} &\qw & \qw & \ctrl{1} & \qw & \qw  &\qw & \qw  & \ctrl{1}    &\qw \\
& \gate{H} & \targ  &\qw  & \ctrl{1} & \qw & \ctrl{1} & \targ & \gate{H^{\dagger}} &\qw & \gate{S} & \targ  &\qw  & \ctrl{1} & \qw & \ctrl{1} & \targ & \gate{S^{\dagger}} &\qw & \qw & \targ  &\qw  & \ctrl{1} & \qw & \ctrl{1} & \targ  &\qw\\ 
& \qw &\qw  &\qw &\targ  &\gate{R_z(-c_{1(4)})}  &\targ &\qw & \qw &\qw & \qw &\qw  &\qw &\targ  &\gate{R_z(c_{2(5)})}  &\targ &\qw & \qw &\qw & \qw &\qw  &\qw &\targ  &\gate{R_z(-c_{3(6)})}  &\targ &\qw  &\qw 
}
\vspace{0.5cm}

(b) \quad \Qcircuit @C=0.4em @R=0.7em {
& \gate{H} & \ctrl{1} & \qw & \qw  &\qw & \qw  & \ctrl{1}   & \gate{H^{\dagger}} &\qw & \gate{S} & \ctrl{1} & \qw & \qw  &\qw & \qw  & \ctrl{1}  & \gate{S^{\dagger}} &\qw & \qw & \ctrl{1} & \qw & \qw  &\qw & \qw  & \ctrl{1}    &\qw   &\qw & \qw &\qw & \qw \\
& \gate{H} & \targ  &\qw  & \ctrl{1} & \qw & \ctrl{1} & \targ & \gate{H^{\dagger}} &\qw & \gate{S} & \targ  &\qw  & \ctrl{1} & \qw & \ctrl{1} & \targ & \gate{S^{\dagger}} &\qw & \qw & \targ  &\qw  & \ctrl{1} & \qw & \ctrl{1} & \targ  &\qw  &\qw  &\qw & \qw & \qw \\ 
& \gate{H} &\qw  &\qw &\targ  &\gate{R_z(-a_1)}  &\targ &\qw & \gate{H^{\dagger}} &\qw & \gate{H} &\qw  &\qw &\targ  &\gate{R_z(-a_{2})}  &\targ &\qw & \gate{H^{\dagger}} &\qw  &\gate{H} &\qw  &\qw &\targ  &\gate{R_z(-a_{3})}  &\targ &\qw  &\gate{H^{\dagger}} & \qw &\gate{R_z(-a_{4})} &\qw &\gate{H} &\qw
}
\caption{\label{fig:circuits2} Circuits for (a) $F^{(1(2))}$ and (b) $J$ for $SU(2^3)$.} 
\end{figure*}
Fig.~\ref{fig:circuits2} indicates that the encoding circuits are made up of $CNOT$ gates and single qubit unitaries which are rotations about $z$ axis on the Bloch sphere by angles $\alpha_{i}$ determined by the search parameters.  In other words, once we obtain the optimal code, we can easily encode into the desired subspace by only changing the rotation angle about the $z$ axis, while keeping the rest of the components in the encoding circuit fixed.

\section{{Optimal codes for the amplitude-damping channel}}\label{app:numericalcodes}

We list {in Table \ref{tab1} }the optimal codes obtained using our numerical search, for the standard amplitude-damping channel, corresponding to the plots in Fig.~\ref{fig:3-4qubit}.

\begin{table*}
\begin{tabular}{c||c|c|c||c|c|c}
Code&Encoding&$|0_L\rangle$&$|1_L\rangle$&Encoding&$|0_L\rangle$&$|1_L\rangle$\\
\hline\hline
&&&&&&\\
3-qubit & unstructured & 
~$\begin{pmatrix}
  -0.426 + 0.235\upi\\
   0.040 - 0.415\upi\\
   0.014 + 0.084\upi\\
  -0.312 + 0.323\upi\\
   0.021 + 0.278\upi\\
   0.089 + 0.167\upi\\
  -0.303 + 0.038\upi\\
  -0.403 + 0.108\upi\\
  \end{pmatrix}$~& 
  ~$\begin{pmatrix}
   0.275 + 0.103\upi\\
   0.248 + 0.191\upi\\
   0.116 - 0.116\upi\\
   0.008 - 0.194\upi\\
   0.429 + 0.266\upi\\
  -0.066 - 0.269\upi\\
  -0.086 + 0.305\upi\\
  -0.488 - 0.285\upi\\
  \end{pmatrix}$~& structured & 
~$\begin{pmatrix}
  -0.013 + 0.076\upi\\
  -0.587 + 0.370\upi\\
   0.000 + 0.000\upi\\
   0.000 + 0.000\upi\\
   0.000 + 0.000\upi\\
   0.000 + 0.000\upi\\
   0.026 + 0.052\upi\\
   0.385 + 0.601\upi\\
  \end{pmatrix}$~&
  ~$\begin{pmatrix}
   0.000 + 0.000\upi\\
   0.000 + 0.000\upi\\
  -0.152 + 0.056\upi\\
  -0.330 - 0.177\upi\\
   0.491 + 0.763\upi\\
  -0.044 - 0.095\upi\\
   0.000 + 0.000\upi\\
   0.000 + 0.000\upi\\
  \end{pmatrix}$~\\
&&&&&&\\
\hline
&&&&&&\\
4-qubit & unstructured & 
$\begin{pmatrix}
  0.448 + 0.236\upi\\
  -0.066 + 0.134\upi\\
  -0.052 + 0.003\upi\\
  -0.044 - 0.027\upi\\
  -0.037 + 0.058\upi\\
   0.313 + 0.048\upi\\
  -0.338 - 0.057\upi\\
   0.001 - 0.060\upi\\
   0.006 - 0.114\upi\\
  -0.310 - 0.088\upi\\
  -0.356 - 0.073\upi\\
   0.020 - 0.004\upi\\
   0.059 - 0.004\upi\\
   0.041 - 0.002\upi\\
  -0.038 + 0.040\upi\\
   0.412 + 0.250\upi
\end{pmatrix}$&
$\begin{pmatrix}
   0.379 - 0.350\upi\\
  -0.012 + 0.001\upi\\
  -0.040 + 0.042\upi\\
   0.027 - 0.038\upi\\
   0.038 - 0.024\upi\\
  -0.170 + 0.292\upi\\
   0.191 - 0.321\upi\\
   0.027 - 0.053\upi\\
   0.031 - 0.041\upi\\
   0.200 - 0.276\upi\\
   0.210 - 0.290\upi\\
   0.027 + 0.077\upi\\
  -0.014 + 0.033\upi\\
   0.098 + 0.013\upi\\
   0.022 + 0.026\upi\\
   0.356 - 0.276\upi
\end{pmatrix}$& structured & 
$\begin{pmatrix}
   0.580 - 0.352\upi\\
   0.026 - 0.210\upi\\
   0.027 + 0.040\upi\\
  -0.001 + 0.042\upi\\
   0.000 + 0.000\upi\\
   0.000 + 0.000\upi\\
   0.000 + 0.000\upi\\
   0.000 + 0.000\upi\\
   0.000 + 0.000\upi\\
   0.000 + 0.000\upi\\
   0.000 + 0.000\upi\\
   0.000 + 0.000\upi\\
  -0.014 + 0.030\upi\\
  -0.056 + 0.025\upi\\
   0.134 - 0.166\upi\\
   0.048 + 0.662\upi
\end{pmatrix}$ &
$\begin{pmatrix}
 0.000 + 0.000\upi\\
   0.000 + 0.000\upi\\
   0.000 + 0.000\upi\\
   0.000 + 0.000\upi\\
   0.186 + 0.028\upi\\
  -0.353 + 0.178\upi\\
  -0.434 - 0.017\upi\\
  -0.099 + 0.059\upi\\
  -0.191 + 0.123\upi\\
   0.071 - 0.511\upi\\
  -0.346 + 0.379\upi\\
   0.051 + 0.157\upi\\
   0.000 + 0.000\upi\\
   0.000 + 0.000\upi\\
   0.000 + 0.000\upi\\
   0.000 + 0.000\upi
\end{pmatrix}$\\
&&&&&&
\end{tabular}
\caption{\label{tab1} Optimal codes for amplitude-damping channel plotted in Fig.~\ref{fig:3-4qubit}.}
\end{table*}
\end{document}